\documentclass[prb,preprint]{revtex4-1} 

\usepackage{amsmath}  
\usepackage{amsfonts} 
\usepackage{graphicx} 
\usepackage{color}
\newcommand\bb[1] {   \mbox{\boldmath{$#1$}}  }

\newcommand\del{\bb{\nabla}}

\def\gtsima{$\; \buildrel > \over \sim \;$}
\def\gtsim{\lower.5ex\hbox{\gtsima}}

\newcommand\bcdot{\bb{\cdot}}

\newcommand\dd{\partial}

\newcommand\hmn { {\bar h_{\mu\nu}} }
\newcommand\Tmn { {T_{\mu\nu}} }
\newcommand\Smn { {\cal {S_{\mu\nu}} }}

\newcommand\kk{ \kappa }
\newcommand\beq{ \begin{equation} }
\newcommand\eeq{ \end{equation} }

\begin{document}
\noindent {\it \small Essay written for the Gravity Research Foundation 2021 Awards for Essays on Gravitation. \hfill}

\title{A Poynting theorem formulation for the gravitational wave stress pseudo tensor}

\author{Steven A.\ Balbus}
\affiliation{Department of Physics, Astrophysics, University of Oxford, Denys Wilkinson Building, Keble Road, Oxford OX1 3RH, United Kingdom\footnote{\large email: steven.balbus@physics.ox.ac.uk}}



\date{\today}

\begin{abstract}
\baselineskip=18pt
A very simple and physical derivation of the conservation equation for the propagation of gravitational radiation is presented.   The formulation is exact.     The result takes the readily recognisable and intuitive form of a Poynting-style equation, in which the outward propagation of stress-energy is directly related to the volumetric equivalent of a radiation reaction force acting back upon the sources, including the purely gravitational contribution to the sources.     
Upon averaging, the emergent pseudo tensor for the gravitational radiation is in exact agreement with that found by much more labour-intensive methods.   
 
\end{abstract}

\maketitle 

\section{Introduction} 

\baselineskip = 18pt

Whilst there is no unique expression for {\em the} stress energy tensor for gravitational radiation,  the energy actually carried off to infinity by gravitational waves is well-defined and can be determined unambiguously\cite{note, mis}.     Deriving an analogue of the stress energy tensor for this radiation, often referred to as a ``pseudo tensor,'' is a famously nuanced and rather cumbersome exercise at both the conceptual and technical levels.   An oft-used approach\cite{wein} is to work in quasi-Cartesian coordinates, writing the metric tensor $g_{\mu\nu}$ as the sum\cite{note}
\beq
g_{\mu\nu} = \eta_{\mu\nu} + h_{\mu\nu},
\eeq
where $\eta_{\mu\nu}$ is the standard Minkowski metric and $h_{\mu\nu}$ the departure therefrom, not necessarily small, except by assumption at large distances.    The classic Einstein tensor,
\beq
G_{\mu\nu} \equiv R_{\mu\nu} - {g_{\mu\nu}R\over 2},
\eeq
where $R_{\mu\nu}$ is the Ricci tensor and $R$ its trace the Ricci scalar, is then expanded in powers of the amplitudes of $h_{\mu\nu}$ and its derivatives.     This expansion is then substituted into the Einstein field equations  
\beq
G_{\mu\nu} = - 8\pi GT_{\mu\nu},
\eeq
and defining
\beq 
t_{\mu\nu} = \sum_{n=2}^\infty G^{(n)}_{\mu\nu},
\eeq
where $G^{(n)}_{\mu\nu}$ denotes the successive terms of $G_{\mu\nu}$ of multiplicative order $n$ in the  $h_{\mu\nu}$ amplitudes, the exact nonlinear field equations become
\beq\label{G1}
G^{(1)}_{\mu\nu} = - 8\pi G (T_{\mu\nu} +t_{\mu\nu}).
\eeq
To leading quadratic order, the tensor $t_{\mu\nu}$  is then identified as the stress energy of the gravitational radiation itself.   After a rather cumbersome calculation, one finds  
\beq\label{tt}
t_{\mu\nu} = {1\over 32\pi G}\Big[ \left\langle 
{\dd_\mu \bar h_{\kappa\lambda}} \,  {\dd_\nu{\bar h}^{\kappa\lambda}}   \right\rangle- \left\langle {\dd_\lambda {\bar h}^{\lambda\kappa}}  \, {\dd_\nu {\bar h}_{\kappa\mu}}\right\rangle- \left\langle {\dd_\lambda {\bar h}^{\lambda\kappa}} \, {\dd_\mu {\bar h}_{\kappa\nu}}\right\rangle -{1\over 2}\left\langle{\dd_\mu \bar h}\,
{\dd_\nu \bar h}\right\rangle\Big],
\eeq
where
$$
\hmn = h_{\mu\nu} - {\eta_{\mu\nu}\over 2}h, \quad h\equiv h^\mu_{\ \mu}, \quad \bar h \equiv \bar h^\mu_{\ \mu}=-h.
$$
This form of $t_{\mu\nu}$, which has been spatially averaged, is gauge-invariant\cite{mis}.   Equation (\ref{tt}) and its gauge invariance emerge only in this averaged sense, which is taken over many effective periods of the wave, so that the purely oscillatory quadratic terms are ignored.   (These generally take the form of exact divergences.)     The averaging is indicated by the angle bracket $\langle\rangle$.     Considerable simplification is then possible with a prudent choice of gauge.    The ``harmonic gauge,'' analogous to the Lorenz gauge of electromagnetic theory, is a convenient choice for the study gravitational waves, as it eliminates many terms in the initial expansion from which $t_{\mu\nu}$ is distilled.  If $h_{\mu\nu}$  has a plane wave dependence of the form $\exp(ik_\mu x^\mu)$,  a harmonic gauge must be used if a standard null $k_\mu$ 
($k_\mu k^\mu=0$) is assumed.    All physical, curvature-inducing gravitational radiation have this property\cite{mis}, as opposed to merely being ``sinuosities in the co-ordinate system,  and the only speed of propagation relevant to them is the speed of thought\cite{edd}.''   The harmonic gauge is defined by 
\beq
\dd_\mu\bar h^{\mu\nu} = 0 \qquad {\rm (harmonic\ gauge\ condition).}
\eeq
It is always possible to construct such a gauge \cite{mis} by solving a classical wave equation, just as it is always possible to work in the Lorenz gauge in electrodynamics by the same route.  
In the ``transverse traceless'' (TT) gauge, which is itself harmonic, there is an additional constraint $h=0$, which leads to the simple result
\beq\label{ttg}
t_{\mu\nu}={1\over 32\pi G}\left\langle 
{\dd_\mu \bar h_{\kappa\lambda}} \   {\dd_\nu{\bar h}^{\kappa\lambda}}   \right\rangle\quad{\rm (TT\ gauge).}
\eeq

One may ask whether this very roundabout route for arriving at the dynamical properties of gravity waves is one that is simply unavoidable.   It has been shown \cite{bal}, in fact, that in the context of strictly linear theory, the expression (\ref{tt}) does indeed emerge much more straightforwardly, as a simple moment of the linear wave equation.    This avoids the need of abstracting explicit second order terms from the geometrical Einstein tensor, designating them with special status, and then undertaking an lengthy ``weeding'' operation  to keep only a few essential terms.    Here we shall show that the methods that have been used successfully in linear theory may be generalised to a fully nonlinear theory.  This allows for a Poynting-like formulation of the problem, in which the work done by gravitational field acts inclusively upon its energy sources of both a material and gravitational nature.    It all comes down to taking the first moment of a wave equation.

\section{Analysis}

\subsection {Conserved stress energy  flux}

Equation (\ref{G1}), written explicitly on the left-hand side, is \cite{mis}:
\beq\label{1}
\Box \hmn- {\dd_\nu\dd_\lambda{\bar h}^\lambda_\mu} - {\dd_\mu\dd_\lambda {\bar h}^\lambda_\nu} +\eta_{\mu\nu} {\dd_\lambda\dd_\rho{\bar h}^{\lambda\rho} }=-\kk (\Tmn+ t_{\mu\nu}) \equiv -\kk\tau_{\mu\nu},
\eeq
where $\Box =\dd_\rho\dd^\rho$, $\kk=16\pi G$, and we have implicitly defined $\tau_{\mu\nu}$ as the pseudo tensor for the sum of the standard material source tensor and the contributions of the ``gravitational field'' itself.    Our coordinates are quasi-Cartesian, and $\bar h_{\mu\nu}$ is not assumed to be small, except asymptotically so at infinity.   We raise (or lower) lower indices on $\dd_\mu$,  $\tau_{\mu\nu}$ and $\bar h_{\mu\nu}$ with $\eta^{\mu\nu}$ ($\eta_{\mu\nu}$).   With this convention, note that
\beq
\dd_\nu\tau^{\mu\nu} = 0
\eeq
is satisfied exactly, since the ordinary divergence of the left side of (\ref{1}) vanishes identically as a simple manifestation of the Bianchi Identities. 

Contracting equation (\ref{1}) with $\eta^{\mu\nu}$, we find:
$$
\Box\bar h + 2 {\dd_\lambda\dd_\rho{\bar h}^{\lambda\rho} }=-\kk \eta^{\mu\nu} \tau_{\mu\nu}\equiv -\kk \tau.
$$
Therefore,
\beq
{\dd_\lambda\dd_\rho {\bar h}^{\lambda\rho} }= -(\Box\bar h +\kk \tau)/2,
\eeq
and our original equation becomes
\beq\label{eq}
\Box \hmn- {\dd_\nu \dd_\lambda{\bar h}^\lambda_\mu} - {\dd_\mu\dd_\lambda {\bar h}^\lambda_\nu} -{\eta_{\mu\nu}\over 2}\Box\bar h =-\kk \Smn,
\eeq
where the (pseudo) source function $\Smn$ is
\beq
\Smn = \tau_{\mu\nu} -{{\eta}_{\mu\nu}\tau\over 2}.
\eeq

Next, contract (\ref{eq}) with $\dd_\sigma \bar h^{\mu\nu}$.    After integrating each of the four terms on the left by parts, the equation takes on the exact and compact form:
\beq\label{gov}
\dd^{\rho} \overline{\cal T}_{\rho\sigma} = -{1\over 2}\  (\dd_\sigma\bar h^{\mu\nu})  \Smn  = -{1\over 2}\  (\dd_\sigma h^{\mu\nu})  \tau_{\mu\nu},
\eeq
where we have defined
\beq  
\overline{\cal T}_{\rho\sigma}  = {\cal T}_{\rho\sigma}  -{\eta_{\rho\sigma}\over 2} {\cal T},
\eeq
and ${\cal T}_{\rho\sigma}$ is a flux pseudo tensor:
\beq\label{T}
{\cal T}_{\rho\sigma} ={1\over \kk} \left[  {1\over 2}\left( 
{\dd_\rho\hmn} \   {\dd_\sigma{\bar h}^{\mu\nu}}   \right)- \left( {\dd_\lambda {\bar h}^{\lambda\mu}}\ {\dd_\sigma {\bar h}_{\mu\rho}}\right)- {1\over 4}\left({\dd_\rho\bar h} \ 
{\dd_\sigma \bar h}\right)\right].
\eeq
with ${\cal T} = \eta^{\mu\nu} {\cal T}_{\mu\nu}$.   Equation (\ref{gov}) is exact, having undergone neither linearisation nor wave averaging.   Note the important reappearance of $h_{\mu\nu}$ and $\tau_{\mu\nu}$ on the right side of the equation.    Note as well that ${\cal T}_{\rho\sigma}$ is very nearly the pseudo tensor $t_{\rho\sigma}$ of equation (\ref{tt}).   The latter has been averaged and is manifestly symmetric, whereas
equation (\ref{T}) is not symmetric in its indices due to the second term.    We shall assume that the ${\bar h}_{\mu\nu}$ fall off as $1/r$ at large $r$ so that the corresponding stress energy flux passing through a sphere of infinite
radius remains finite.    This seems physically reasonable (and certainly in accord with LIGO observations), but it also constrains the partner integral over all space on the right side of equation (\ref{gov}) to be convergent, despite the distributed nature of its wave source $\tau_{\mu\nu}$.  

That the anomalous second term of ${\cal T}_{\rho\sigma}$ vanishes in a harmonic gauge, leaving a manifestly symmetric pseudo tensor, suggests that the asymmetry is apparent, not real.    In fact, if at this point we introduce the standard gauge transformation 
\beq
\bar h_{\mu\nu} \rightarrow   \bar h_{\mu\nu} -\dd_\mu \xi_\nu - \dd_\nu \xi_\mu,
\eeq
where the $\xi_\mu$ are arbitrary functions, the gauge invariance of $\langle{\cal T}_{\rho\sigma}\rangle$ may be readily established, index asymmetry and all, once the standard averaging procedure is applied.    (Recall that the averaging relies on ignoring exact derivatives when integrating by parts, which may be justified for the far-field waves of interest.)

There follow two immediate consequences.    First, since the trace $\langle{\cal T}\rangle$ must also be gauge invariant, and it clearly vanishes in a harmonic gauge with $k^\rho k_\rho=0$, it therefore vanishes in any gauge and may henceforth be ignored.  (More strictly, in view of the background curvature from the waves themselves, the $k^\rho k_\rho$ terms are {\em higher order} in smallness and do not contribute to the distant outward flux.)    Second, since the far-field ${\langle\cal T}_{\rho\sigma}\rangle$ is manifestly symmetric in a harmonic gauge and is likewise gauge invariant, it may be replaced by its symmetrised form:
\beq\label{Tavg}
\langle {\cal T}_{\rho\sigma}\rangle = {1\over 2\kk}\left[
\langle {\dd_\rho\hmn} \   {\dd_\sigma{\bar h}^{\mu\nu}}\rangle -
 \langle {\dd_\lambda {\bar h}^{\lambda\mu}}\ {\dd_\sigma {\bar h}_{\mu\rho}}\rangle-
\langle {\dd_\lambda {\bar h}^{\lambda\mu}}\ {\dd_\rho {\bar h}_{\mu\sigma}}\rangle
- {1\over 2}\langle{\dd_\rho\bar h} \  {\dd_\sigma \bar h}\rangle,
\right]
\eeq
which now is precisely equation (\ref{ttg}).    Under time-steady conditions, stress energy conservation may be expressed globally as:
\beq\label{Poynt}
\int_\infty \langle {\cal T}_{\sigma i}\rangle n_i \, dS = -{1\over 2}  \int \tau_{\mu\nu}    \dd_\sigma h^{\mu\nu}\, d^3 r,
\eeq
where the integral on the left is over a spherical surface at spatial infinity.   Energy conservation corresponds to $\sigma =0$.    Note the use of Euclidian geometry on both sides of the equation, and the implicit assumption, discussed earlier, that the integral on the right converges.      


\subsection {Wave generation via energy loss}

The attentive reader will note that we have not yet justified the overall normalisation of $\langle{\cal T}_{\rho\sigma}\rangle$.  This, and more, is contained in the right side of equation (\ref{gov}).   We first set $\sigma=0$ in order to focus on the energy flux $\langle{\cal T}_{\rho0}\rangle$.   Then, to establish the normalisation, we take the quasi-Newtonian limit\cite{bal} of the source term of (\ref{gov}), and integrate by parts ignoring the divergence terms (a process denoted by ``$\rightarrow$''):{\small 
\beq
 -{1\over 2}\  (\dd_0 h^{00})  T_{00} =  -{1\over 2}\  (\dd_0 h^{00})  T^{00} \rightarrow {1\over 2}\  h^{00}  \dd_0 T^{00}  = -{1\over 2}\  h^{00}  \dd_i T^{0i} \rightarrow  {1\over 2}\  \dd_i h^{00} T^{0i}\simeq - \rho\bb{v}\bcdot \del\Phi.
\eeq}
Here, we have introduced the spatial index $i$, rest mass density $\rho$, ordinary velocity vector $\bb{v}$ and gravitational potential function $\Phi$.    Clearly, the right side reduces to a properly normalised ``work done on the sources'' expression in the limit of lowest nonvanishing order; the sign is correct when the left side of the equation is written in covariant form for $\langle {\cal T}_{\rho\sigma}\rangle$.     The actual potential function that does the work on the sources in this limit is the radiation-reaction, Burke-Thorne potential\cite{bur,thor}, accessible in a suitable gauge\cite{mis},  for which the fully gauge-invariant form of our governing equation is required.         

Finally, we may note that 
\beq 
-{1\over 2}\  (\dd_0 h^{\mu\nu})  \tau_{\mu\nu}
\eeq
is the precise generalisation of ``work done'' by radiation.   In particular, $\tau_{\mu\nu}$ includes in its pseudo tensorial manner the stress energy of the gravitational field itself.    In other words, this remarkable expression uses the radiation reaction component of the $h_{\mu\nu}$ gradient to act back on the stress energy of the those very waves associated with the graviational radiation that are escaping to infinity!    This is analogous to the well-known observation that the integration constant $M$ of either a Schwarzschild or Kerr black hole embodies not only the conventional mass-energy that has gone into the creation of the hole, but a gravitational contribution as well.  

We may define an angular momentum counterpart to $\cal T_{\rho\sigma}$:
\beq
{\cal J}_{i\rho} =   \epsilon_{ijk} x_j \langle {\cal T}_{\rho k} \rangle= x_j \langle {\cal T}_{\rho k}\rangle - x_k \langle {\cal T}_{\rho j} \rangle
\eeq
where $\epsilon_{ijk}$ is the Levi-Civita symbol, which allows a normal-ordering of the distinct spatial indices $i,j,k$. Placement of the spatial indices carries no physical significance in these coordinates.  The final equality clearly allows for the definition of a more general spacetime third-order pseudo tensor,  but we shall have no need to avail ourselves of this here.   For present purposes, it will suffice to note that equation (\ref{gov}) and symmetry of $\langle {\cal T}_{\rho k}\rangle$,  lead directly to an equation of global angular momentum conservation:
\beq\label{angm}
\int_\infty {\cal J}_{i j}\, n_j dS  =  -{1\over 2}\epsilon _{ijk}\int x_j (\dd_k h^{\mu\nu})  \tau_{\mu\nu}\, d^3 r
\eeq
where the integrals are Euclidian as in equation ({\ref{Poynt}).  
Once again we may note that the stress energy of the gravitational waves plays an explicit role as a contributor to its own angular momentum source.



%

\section {Closing thoughts}
Lighthill\cite{lthl} has made the point that in the context of waves in fluids, whilst transport involves second-order products in the wave amplitudes, it is generally not necessary to perform second-order calculations to obtain the energy and angular momentum fluxes.   It is therefore gratifying that the equations of gravitational wave propagation prove not to be an exception to Lighthill's dictum.    There is no more information in the wave stress-energy pseudo tensor than there is in the linearised wave equation itself, any more than an equation of mechanical energy conservation embodies more information than Newton's Second Law of motion.    There is however, a certain satisfaction to be had in knowing that buried within the core of the extensive $G^{(2)}_{\mu\nu}$ component of the field equations is a simple first moment of $G^{(1)}_{\mu\nu}$.     

There is also the satisfaction of being able to see directly {\em why} $t_{\mu\nu}$ has the form it does in the first place, rather than having it just emerge at the end of a lengthy distillation process.    Indeed, if we apply our moment method to the simple scalar field equation $\Box \Phi=4\pi G\rho$, where $\Phi$ is the potential and $\rho$ the source density, we are led forthwith
to a wave energy flux of $-(\dd_t\Phi)(\dd_i\Phi)/4\pi G$.     On the other hand, we have also seen that the true gravitational wave energy flux in the TT gauge is given by $-(\dd_t h^{jk})(\dd_i h_{jk})/32 \pi G$ which, for a single mode of wave polarisation (say the $+$ mode), is $-(\dd_t h^{11})(\dd_i h_{11})/16 \pi G$.   If we now brazenly identify the $h^{11}$ amplitude with a $-2\Phi$ potential counterpart, we are led directly back to the result of our simple scalar theory.    Hardly a rigorous argument!     But the reader may agree that it is not without a certain suggestive charm, and if we recall in particular how the overall normalisation of the field equations is achieved via the invocation of a Newtonian limit, perhaps not entirely devoid of content.   

Finally, let us recall the elegant covariant formulation of Poynting's theorem from Maxwellian electrodynamics:
\beq
\dd_\alpha T^{\alpha\beta}_{\rm em} = - J^\beta F^\alpha_{\  \beta}
\eeq
where $T^{\alpha\beta}_{\rm em} $ is the electromagnetic stress energy tensor, $J^\beta$ the 4-current, and $F^{\alpha}_{\ \beta}$ the mixed form of the electromagnetic field tensor\cite{jac, wein}.    That the field equations of gravitation also allow such an analogous formulation via equation (\ref{gov}), with the ``gravitational field''  serving as its own source,  is both enlightening and instructive.    If one forgoes the need for a manifestly symmetric $\overline{\cal T}_{\rho\sigma}$ this can be achieved formally with no averaging at all, but to elicit the true physical content, especially with regard to angular momentum conservation, requires an averaging procedure, and a statement of conservation best expressed globally via equations (\ref{Poynt}) or (\ref{angm}).     We physicists are most fortunate to be living in a time when the detection of gravitational radiation and its use as a diagnostic in astronomy are becoming routine.   It is more timely than ever to be able to understand these fascinating ripples in spacetime (or, for TT gauge aficionados, in space alone) on terms that are as familiar and physically motivated as possible.

\begin{acknowledgments}

I am grateful to P.\ Ferreira, L.\ Fraser-Taliente, and A.\ Mummery for their encouraging and constructive comments on an earlier draft of this work.    I am pleased to acknowledge support from the Hintze Family Charitable Foundation and STFC (grant ST/S000488/1) for my research.      


\end{acknowledgments}


\begin{thebibliography}{99}
\bibitem{note} 
In this paper, Greek indices indicate spatio-temporal dimensions; Roman
indices spatial dimensions.   The speed of light $c=1$.  We use the sign convention $[-+++]$ for the diagonal Minkowski metric $\eta_{\mu\nu}$. 
\end{thebibliography}
\end{document}